\documentclass{PoS}

\usepackage{graphicx}

\usepackage{wrapfig,float,slashed}
\usepackage{amsmath,dsfont}
\usepackage{amssymb,epsfig,graphicx}
\usepackage[table]{xcolor}
\usepackage{epstopdf}
\usepackage[normalem]{ulem}
\usepackage{ragged2e}
\usepackage{lipsum}
\usepackage[font=small,skip=1pt]{caption}
\allowdisplaybreaks
\usepackage{pifont}

\usepackage[hang,flushmargin,multiple]{footmisc}

\usepackage{tikz}
\usetikzlibrary{arrows,positioning,shapes,mindmap}
\usetikzlibrary{decorations.pathmorphing}
\usetikzlibrary{decorations.markings}

\newcommand{\nc}{\newcommand}
\nc{\non}{\nonumber}
\nc{\hc}{\hbox {H.c.}}
\nc{\noi}{\noindent}
\nc{\barx}{\bar{x}}
\nc{\pbarn}{\;\hbox {pb}}
\nc{\fbarn}{\;\hbox {fb}}
\nc{\hsp}{\hspace{0.5cm}}
\nc{\lsp}{\hspace{1cm}}
\nc{\Lsp}{\hspace{2cm}}
\nc{\LLsp}{\lsp\lsp}
\nc{\lra}{\longrightarrow}
\nc{\p}{\prime}
\nc{\sgn}{\text{sgn}}
\nc{\arccot}{\text{arccot}}
\nc{\ph}{\varphi}
\nc{\co}{{\cal O}}
\nc{\beq}{\begin{equation}}  \nc{\eeq}{\end{equation}}
\nc{\bea}{\begin{eqnarray}}  \nc{\eea}{\end{eqnarray}}
\nc{\baa}{\begin{array}}     \nc{\eaa}{\end{array}}
\nc{\bit}{\begin{itemize}}   \nc{\eit}{\end{itemize}}
\nc{\ben}{\begin{enumerate}} \nc{\een}{\end{enumerate}}
\nc{\bce}{\begin{center}}    \nc{\ece}{\end{center}}
\nc{\bpm}{\begin{pmatrix}}   \nc{\epm}{\end{pmatrix}}
\nc{\bvt}{\begin{verbatim}}  \nc{\evt}{\end{verbatim}}

\def\lsim{\mathrel{\raise.3ex\hbox{$<$\kern-.75em\lower1ex\hbox{$\sim$}}}}
\def\gsim{\mathrel{\raise.3ex\hbox{$>$\kern-.75em\lower1ex\hbox{$\sim$}}}}

\def\udots{\mathinner{\mkern1mu\raise1pt\vbox{\kern7pt\hbox{.}}\mkern2mu\raise4pt\hbox{.}\mkern2mu\raise7pt\hbox{.}\mkern1mu}}

\def\gev{\;\hbox{GeV}}
\def\tev{\;\hbox{TeV}}

\def\mst{M_\ast}
\definecolor{agray}{rgb}{0.95, 0.95, 0.99}

\newcommand\fverb{\setbox\fverbbox=\hbox\bgroup\verb}
\newcommand\fverbdo{\egroup\medskip\noindent%
			\fbox{\unhbox\fverbbox}\ }
\newcommand\fverbit{\egroup\item[\fbox{\unhbox\fverbbox}]}
\newbox\fverbbox

\makeatletter
\renewcommand{\boxed}[1]{\textcolor{black}{%
\tikz[baseline={([yshift=-0ex]current bounding box.center)}] \node [rectangle, minimum width=0ex,draw] {\normalcolor\m@th$\displaystyle#1$};}}
 \makeatother

\title{Higgs dark matter from a warped extra dimension}

\ShortTitle{Higgs dark matter from a warped extra dimension}

\author{\speaker{Aqeel Ahmed}\\
Faculty of Physics,
University of Warsaw,
Pasteura 5, 02-093 Warsaw, Poland \\
National Centre for Physics, Quaid-i-Azam University Campus, Islamabad, Pakistan\\
        E-mail: \email{aqeel.ahmed@fuw.edu.pl}}
\author{Bohdan Grzadkowski\\
        Faculty of Physics,
University of Warsaw,
Pasteura 5, 02-093 Warsaw, Poland\\
        E-mail: \email{bohdan.grzadkowski@fuw.edu.pl}}

\author{John~F.~Gunion\\
        Department of Physics, University of California, Davis, CA 95616, U.S.A.\\
        E-mail: \email{jfgunion@ucdavis.edu}}

\author{Yun Jiang\\
Niels Bohr International Academy, University of Copenhagen,
Blegdamsvej 17, DK-2100
Copenhagen, Denmark\\
        E-mail: \email{yunjiang@ucdavis.edu}}

\abstract{
We present a 5D $\mathbb{Z}_2$-symmetric IR-UV-IR model with a {\it warped KK-parity} under which the bulk fields have towers of either even or odd KK-modes. We show that this $\mathbb{Z}_2$-symmetric geometry is equivalent to two times the UV-IR geometry (Randall-Sundrum model) provided each bulk field is subject to Neumann (or mixed) and Dirichlet boundary conditions at the UV-brane for even and odd fields, respectively. The 5D Standard Model (SM) bosonic sector is considered, such that in the 4D low-energy effective theory the $\mathbb{Z}_2$-even zero-modes correspond to the SM degrees of freedom, whereas the
$\mathbb{Z}_2$-odd zero modes serve as a dark sector.
In the zero-mode scalar sector, the even scalar mimics the SM Higgs boson, while the odd scalar (dark-Higgs) is stable and serves as a dark matter candidate. Implications for this dark matter are discussed; it is found that the dark-Higgs can provide only a small fraction of the observed dark matter abundance.
}

\FullConference{18th International Conference From the Planck Scale to the Electroweak Scale \\
		 25-29 May 2015\\
		 Ioannina, Greece }
		
\begin{document}

\tikzstyle{every picture}+=[remember picture]
\pgfdeclaredecoration{complete sines}{initial}
{
    \state{initial}[
        width=+0pt,
        next state=sine,
        persistent precomputation={\pgfmathsetmacro\matchinglength{
            \pgfdecoratedinputsegmentlength / int(\pgfdecoratedinputsegmentlength/\pgfdecorationsegmentlength)}
            \setlength{\pgfdecorationsegmentlength}{\matchinglength pt}
        }] {}
    \state{sine}[width=\pgfdecorationsegmentlength]{
        \pgfpathsine{\pgfpoint{0.25\pgfdecorationsegmentlength}{0.5\pgfdecorationsegmentamplitude}}
        \pgfpathcosine{\pgfpoint{0.25\pgfdecorationsegmentlength}{-0.5\pgfdecorationsegmentamplitude}}
        \pgfpathsine{\pgfpoint{0.25\pgfdecorationsegmentlength}{-0.5\pgfdecorationsegmentamplitude}}
        \pgfpathcosine{\pgfpoint{0.25\pgfdecorationsegmentlength}{0.5\pgfdecorationsegmentamplitude}}
}
    \state{final}{}
}
\tikzset{
fermion/.style={thick,draw=black, line cap=round, postaction={decorate},
    decoration={markings,mark=at position 0.6 with {\arrow[black]{latex}}}},
photon/.style={thick, line cap=round,decorate, draw=black,
    decoration={complete sines,amplitude=4pt, segment length=6pt}},
boson/.style={thick, line cap=round,decorate, draw=black,
    decoration={complete sines,amplitude=4pt,segment length=8pt}},
gluon/.style={thick,line cap=round, decorate, draw=black,
    decoration={coil,aspect=1,amplitude=3pt, segment length=8pt}},
scalar/.style={dashed, thick,line cap=round, decorate, draw=black},
ghost/.style={dotted, thick,line cap=round, decorate, draw=black},
->-/.style={decoration={
  markings,
  mark=at position 0.6 with {\arrow{>}}},postaction={decorate}}
 }

\makeatletter
\tikzset{
    position/.style args={#1 degrees from #2}{
        at=(#2.#1), anchor=#1+180, shift=(#1:\tikz@node@distance)
    }
}
\makeatother

\section{Introduction}
\label{Introduction}
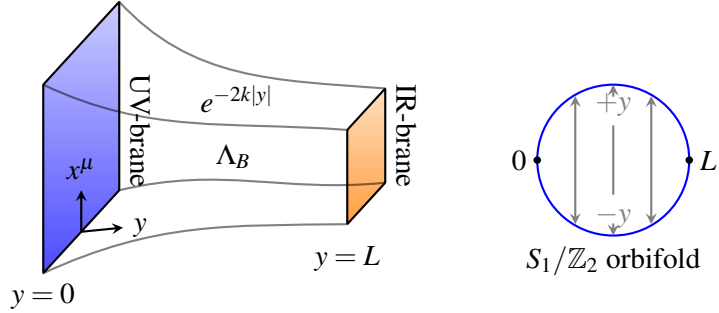
\begin{figure}
\centering
\begin{tikzpicture}[thick,rounded corners=0.5pt,line cap=round,scale=0.5]
\shadedraw[top color=blue!20,bottom color=blue!70,yslant=0.1]
(0,0) -- (2,2) -- node[above=-13pt,rotate=-90]{UV-brane} (2,7) -- (0,5) --  cycle;
\shadedraw[top color=orange!25,bottom color=orange!75,yslant=0.1](8,0.5) -- (9,1.5) -- node[above,rotate=-90]{IR-brane} (9,4) -- (8,3) -- cycle;
\draw[thick,black,yslant=0.1,opacity=0.5](0,5) to [out=-30,in=170]node[above right,opacity=1]{$e^{-2k|y|}$} (8,3)
                                (2,7) to [out=-50,in=170] (9,4)
                                (2,2) to [out=8,in=180] (9,1.5)
                                (0,0) to [out=20,in=175] (8,0.5);
\draw (0,0) node[below]{$y=0$}
 (8,1)node[below]{$y=L$}
 (5,2.5)node[above,black]{$\Lambda_B$};
\draw[->,>=stealth,thick,yslant=0.1] (1,1) -- (1,2.1) node[above]{$x^\mu$};
\draw[->,>=stealth,thick,yslant=0.1](1,1)-- (2.1,1) node[right]{$y$};
\begin{scope}[xshift=15cm,yshift=3cm]
\draw[blue] (0,0)circle (2cm);
\draw[<-,>=stealth,thick,gray] (0,2) -- (0,1.7);
\draw[<-,>=stealth,thick,gray] (0,-2)node[below, black]{$S_1/\mathbb{Z}_2$ orbifold} -- (0,-1.7);
\draw[<->,>=stealth,thick,gray] (1,1.7) -- (1,-1.7);
\draw[<->,>=stealth,thick,gray] (-1,1.7) -- (-1,-1.7);
\draw[thick, gray] (0,0.9)node[above]{$+y$} -- (0,-0.9)node[below]{$-y$};
\filldraw [black] (-2,0)node[left,black]{$0$} circle (2pt);
\filldraw [black] (2,0)node[right,black]{$L$} circle (2pt);
\end{scope}
\end{tikzpicture}
\caption{Cartoon of RS1 geometry.}
\label{fig_rs_geometry}
\end{figure}

The 5D warped model of Randall and Sundrum (RS) with two D3-branes (RS1) provides an elegant solution to {\it the hierarchy problem} \cite{Randall:1999ee}. The two D3-branes are localized at the fixed points of the $S_1/\mathbb{Z}_2$ orbifold, a ``UV-brane'' at $y=0$ and  an ``IR-brane'' at $y=L$ (UV-IR model), see Fig.~\ref{fig_rs_geometry}.
The metric for the RS geometry is \cite{Randall:1999ee},
\beq
ds^2=e^{-2k|y|}\eta_{\mu\nu}dx^\mu dx^\nu+dy^2,  \label{metric}
\eeq
where $k$ is a constant of the order of 5D Planck mass $M_\ast$. Randall and Sundrum showed that if the 5D theory involves only one mass scale
$M_\ast$ then, due to the presence of non-trivial warping along the extra-dimension, the effective mass scale on the IR-brane is rescaled to $m_{KK}\equiv ke^{-kL}\sim\co(\text{TeV})$ for $kL\sim\co(37)$.

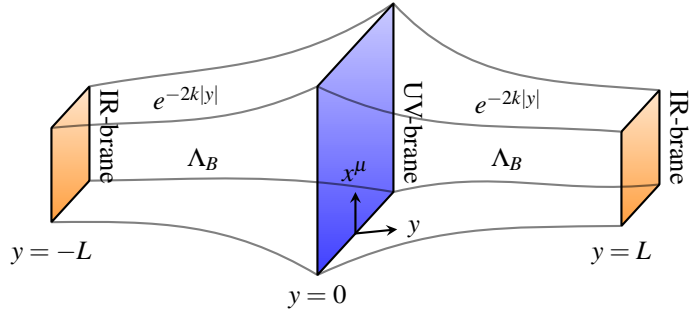
\begin{figure}[t]
\centerline{
\begin{tikzpicture}[thick,rounded corners=0.5pt,line cap=round,scale=0.5]
\shadedraw[bottom color=blue!75,top color=blue!20,yslant=0.1]
(0,0) -- (2,2) -- node[above=-13pt,rotate=-90]{\small UV-brane} (2,7) -- (0,5) --  cycle;
\shadedraw[bottom color=orange!75,top color=orange!20,yslant=0.1](8,0.5) -- (9,1.5) -- node[above,rotate=-90]{\small IR-brane} (9,4) -- (8,3) -- cycle;
\shadedraw[bottom color=orange!75,top color=orange!20,yslant=0.1](-7,2.1) -- (-6,3.1) -- node[above=-1pt,rotate=-90]{\small IR-brane} (-6,5.6) -- (-7,4.6) -- cycle;
\draw[thick,black,yslant=0.1,opacity=0.5](0,5) to [out=-30,in=170]node[above right,opacity=1]{\small $e^{-2k|y|}$} (8,3)
                                (2,7) to [out=-50,in=170] (9,4)
                                (2,2) to [out=8,in=180] (9,1.5)
                                (0,0) to [out=20,in=175] (8,0.5);
\draw[thick,black,yslant=0.1,opacity=0.5](0,5) to [out=200,in=0]node[above,opacity=1]{\small $e^{-2k|y|}$} (-7,4.6)
                                (2,7) to [out=210,in=5] (-6,5.6)
                                (2,2) to [out=165,in=-5] (-6,3.1)
                                (0,0) to [out=145,in=-5] (-7,2.1);
\draw (0,0) node[below]{\small $y=0$}
 (8,0)node[above]{\small $y=L$}
 (-7,0)node[above]{\small $y=-L$}
 (5,2.5)node[above]{\small $\Lambda_B$}
 (-3,2.5)node[above]{\small $\Lambda_B$};
\draw[->,>=stealth,thick,yslant=0.1] (1,1) -- (1,2.1) node[above]{\small $x^\mu$};
\draw[->,>=stealth,thick,yslant=0.1](1,1)-- (2.1,1) node[right]{\small $y$};
\end{tikzpicture}
}
\caption{The geometric configuration for the $\mathbb{Z}_2$-symmetric  IR-UV-IR model.}
\label{IR-UV-IR}
\end{figure}

RS-like warped geometries offer an attractive solution to many of the fundamental puzzles of the SM, mostly through geometric means. In the same spirit, one can ask if RS-like warped extra-dimensions can shed some light on another outstanding puzzle of SM, the lack of a candidate for dark matter (DM) which constitutes 83\% of the observed matter in the universe. It appears that RS1-like models (involving two branes and warped bulk) are unable to offer an analogue of KK-parity, as RS1 geometry is not symmetric around any point along the extra-dimension and hence does not allow a {\it KK-parity}. As a result it cannot accommodate a realistic dark matter candidate. To cure this problem we extend the RS1-like warped geometry in such a way that the whole geometric setup becomes symmetric around a fixed point in the bulk. We construct a IR-UV-IR geometric setup, where
two AdS copies are glued together at the UV fixed point \cite{Ahmed:2015ona}, a cartoon of such a geometric setup is shown in Fig. \ref{IR-UV-IR}. Similar geometric setups are also considered by Refs.~\cite{Agashe:2007jb,Medina:2010mu,Medina:2011qc}.

In this work, we place the SM bosonic sector fields, including the Higgs doublet, in the bulk of the IR-UV-IR geometry. The geometric $\mathbb{Z}_2$ parity ($y\to-y$ symmetry) leads to ``warped KK-parity'', i.e.  there are towers of even and odd KK-modes corresponding to each bulk field. In the weak backreaction scenario we focus on the electroweak symmetry breaking (EWSB) induced by the bulk Higgs doublet and low energy aspects of the 4D effective theory for the even and odd zero-modes assuming the KK-mass scale is high enough $\sim\co(\text{few})\tev$. In the effective theory the even and odd Higgs doublets mimic a two-Higgs-doublet model (2HDM) scenario -- the truncated inert-doublet model -- with the odd doublet similar to the inert doublet but without corresponding pseudoscalar and charged scalars. All the parameters of this  truncated 2HDM are determined by the fundamental 5D parameters of the theory and the choice of boundary conditions (b.c.) for the fields at $y=\pm L$. (Note that the boundary or ``jump'' conditions at $y=0$ follow from the bulk equations of motion in the case of even modes, whereas odd modes are required to be zero by symmetry.) There are many possible alternative choices for the b.c. at $\pm L$. We allow the field to have an arbitrary value at $\pm L$ as opposed to requiring that the field value itself be zero, i.e. we employ Neumann or mixed b.c.  rather than Dirichlet b.c.  at $\pm L$.  With these choices, the symmetric setup yields an odd Higgs zero-mode that is a natural candidate for dark matter. We compute the one-loop quadratic (in cutoff) corrections to the two scalar zero modes within the effective theory and discuss their mass splitting. We calculate the dark matter relic abundance in the cold dark matter paradigm.

The paper is structured as follows. In Sec. \ref{KK-parity from warped geometry}, we provide the background solutions for the IR-UV-IR model and define the warped KK-parity due to the $\mathbb{Z}_2$ geometry. Moreover, in this section we also show that the IR-UV-IR model is equivalent to two times the RS1 geometry (UV-IR),  where each bulk field has an even and an odd field copy, so that each bulk field satisfies the Neumann (or mixed) and Dirichlet b.c. at $y=0$ corresponding to the even and odd KK-modes, respectively. We discuss EWSB for the 5D SM gauge sector due to the bulk Higgs doublet in our ${\mathbb Z}_2$ symmetric model with warped KK-parity in Sec.~\ref{SM EWSB due to bulk Higgs doublet} and obtain a low-energy 4D effective theory containing all the SM fields plus a real scalar -- a dark matter candidate. In subsection~\ref{Dark matter relic abundance} we calculate the relic abundance of the dark-matter candidate. We summarize our findings in Sec. \ref{Summary}.

\section{A ${\mathbb Z}_2$ symmetric IR-UV-IR model and warped KK-parity}
\label{KK-parity from warped geometry}
In this section we provide the background solution for the $\mathbb{Z}_2$ symmetric background (IR-UV-IR) geometry and show how warped KK-parity is manifested within this symmetric warped geometry. We also show that the IR-UV-IR model is equivalent to two times the UV-IR geometry if the bulk fields satisfy both the Neumann (or mixed) and Dirichlet b.c. at the UV-brane, hence providing an even and an odd tower of KK-modes, respectively.

The gravitational action for the $\mathbb{Z}_2$-symmetric IR-UV-IR model can be written as \cite{Ahmed:2015ona},
\beq
S_G=\int d^5x\sqrt{-g}\left\{2\mst^3 R-\Lambda_B-\lambda_{UV}\delta(y)-\lambda_{IR}\big[\delta(y+L)+\delta(y-L)\big]\right\}+S_{GH},
\label{gravity_action}
\eeq
where $R$ is the Ricci scalar, $\Lambda_B$ is the bulk cosmological constant and $\lambda_{UV}(\lambda_{IR})$ is the brane
tension at the UV(IR)-brane. Above and henceforth the Dirac delta functions at end branes are defined in such a way that their integral is 1/2. The action contains the Gibbons-Hawking boundary action,
\beq
S_{GH}=-2\mst^3\int_{\partial{\cal M}}d^4x\sqrt{-\hat g} {\cal K},	\label{GH_term}
\eeq
where ${\cal K}$ is the intrinsic curvature of the surface of the boundary manifold $\partial{\cal M}$.
The solution of the Einstein equations resulting from the above action is the RS metric \eqref{metric}, where the bulk cosmological constant $\Lambda_B$ is related to the brane tensions as \cite{Ahmed:2015ona}
\beq
\lambda_{UV}=- \lambda_{IR}=24\mst^3k, \Lsp
k\equiv\sqrt{\frac{-\Lambda_B}{24\mst^3}},
\eeq
which implies that one needs a positive tension brane at $y=0$ and two negative tension branes at $y=\pm L$.

We would like to comment here that the size of extra dimension could be stabilized in the IR-UV-IR setup through a Goldberger and Wise (GW) mechanism \cite{Goldberger:1999uk,DeWolfe:1999cp}. One of our aims is to analyse EWSB due to a 5D $SU(2)$ Higgs doublet in the IR-UV-IR model, it turns out that one can also employ the same bulk $SU(2)$ Higgs doublet as the GW stabilizing field, see e.g. Refs. \cite{Ahmed:2015ona,Ahmed:2015us,Ahmed:2015prep,Geller:2013cfa}.

\subsection{Warped KK-Parity}
\label{KK-Parity}
We assume that the geometric $\mathbb{Z}_2$-symmetry considered above is exact for our 5D theory. If the 5D theory has this $\mathbb{Z}_2$-parity (symmetry) then the Schr\"odinger-like potential for the bulk fields is symmetric, hence all the eigenmodes (wave-functions) of the Schr\"odinger-like equation are either even (symmetric) or odd (antisymmetric). A general field $\Phi(x,y)$ can be KK decomposed
as
\beq
\Phi(x,y)=\sum_{n}\phi_n(x)f_n(y).
\eeq
Due to the $\mathbb{Z}_2$ geometry, the wave functions $f_n(y)$ are either even or odd, so that $\Phi(x,y)$ can be written as
\beq
\Phi(x,y)\equiv\Phi^{(\pm)}(x,y),
\eeq
with
\begin{align}
\Phi^{(+)}(x,y)&=\sum_n\phi^{(+)}_n(x)f^{(+)}_n(y)\xrightarrow{y\to-y}+\Phi^{(+)}(x,y),   \\
\Phi^{(-)}(x,y)&=\sum_n\phi^{(-)}_n(x)f^{(-)}_n(y)\xrightarrow{y\to-y}-\Phi^{(-)}(x,y).
\end{align}
Due to the geometric $\mathbb{Z}_2$ symmetry, an odd number of odd KK-modes cannot couple to an even number of even KK-modes in the 4D effective theory.
Therefore, the lowest odd KK-mode will be stable and may serve as a dark matter candidate.

Our choice of b.c. will be such that the odd (even) modes satisfy Dirichlet (Neumann or mixed)  boundary (jump) conditions (b.c.) at $y=0$, respectively. As for the odd modes, continuity implies that they must be zero at $y=0$. At the IR-branes we choose the Neumann (or mixed) b.c. for both even and odd modes.

\subsection{RS1 relation to the IR-UV-IR model}
\label{RS1 relation to the IR-UV-IR model}
Let us consider the action for a bulk real scalar field $\Phi^{(\pm)}(x,y)$ in the IR-UV-IR model~\footnote{Due to the warped KK-parity each bulk field is even $(+)$ or odd $(-)$ in the IR-UV-IR model.}\footnote{In our notations the capital Roman indices represent five-dimensional (5D) objects, i.e. $M,N,\cdots=0,1,2,3,5$ and the Greek indices label four-dimensional (4D) objects, i.e.
$\mu,\nu,\cdots=0,1,2,3$.},
\begin{align}
S_{\text{IR-UV-IR}}=-\int d^4x\int_{-L}^{L}dy \sqrt{-g}\bigg\{&\frac12 g^{MN}\nabla_M \Phi^{(\pm)}\nabla_N\Phi^{(\pm)}+V(\Phi^{(\pm)})   \notag\\
&+\lambda_{UV}(\Phi^{(\pm)})\delta(y)+\lambda_{IR}(\Phi^{(\pm)})\big[\delta(y+L)+\delta(y-L)\big]\bigg\},    \label{toy_action}
\end{align}
where $V(\Phi^{(\pm)})$ and $\lambda_{UV(IR)}(\Phi^{(\pm)})$ are the bulk and brane-localized potentials, respectively.
The above action has an exact $\mathbb{Z}_2$-geometric symmetry and hence it can be written as two times the UV-IR geometry (RS1)  where each bulk field has an even and an odd field copy, i.e.
\begin{align}
S_{\text{UV-IR}}=-\int d^4x\int_{0}^{L}dy \sqrt{-g}\bigg\{&\frac12 g^{MN}\nabla_M \tilde\Phi^{(\pm)}\nabla_N\tilde\Phi^{(\pm)}+V(\tilde\Phi^{(\pm)}) \notag\\ &+\lambda_{UV}(\tilde\Phi^{(\pm)})\delta(y)+\lambda_{IR}(\tilde\Phi^{(\pm)})\delta(y-L)\bigg\},    \label{toy_action1}
\end{align}
where the fields $\tilde\Phi^{(\pm)}(x,y)$ in the UV-IR geometry are related to fields $\Phi^{(\pm)}(x,y)$ in the full IR-UV-IR geometry as: $\tilde\Phi^{(\pm)}(x,y)\equiv\Phi^{(\pm)}(x,y)/\sqrt2$ (the factor of $1/\sqrt2$ takes into account the fact that the geometric volume in UV-IR geometry is half of the full IR-UV-IR geometry). Therefore, the canonically normalized fields in the UV-IR geometry would need rescaled couplings. Above, the bulk fields would be subject to Neumann (or mixed) and Dirichlet b.c. at the UV-brane, in the case of the even and odd fields, respectively.

\section{Standard Model bosonic sector in the IR-UV-IR model}
\label{SM EWSB due to bulk Higgs doublet}
Here we consider the SM bosonic sector in the bulk of the IR-UV-IR model to study the phenomenological implications of our
symmetric geometry. As discussed in the previous section the physics of the full IR-UV-IR setup can be described completely by two times the RS1 geometry, i.e. UV-IR setup. However in that scenario each bulk field would be to subject to Neumann (or mixed) and Dirichlet boundary conditions at $y=0$ for even and odd fields, respectively. Hence, we consider only a single AdS slice (UV-IR geometry) and require that each field satisfy  the b.c. corresponding to even and odd fields, i.e. each bulk field has an even and an odd field copy. The 5D action for the electroweak sector of the SM in the UV-IR geometry can be written as
\begin{align}
S=-2\int d^4x\int_0^Ldy \sqrt{-g}\bigg\{&\frac14 F^{a{(\pm)}}_{MN}F^{aMN}_{(\pm)}+\frac14 B^{(\pm)}_{MN}B_{(\pm)}^{MN}+\left\vert D_M H^{(\pm)}\right\vert^2 +\mu_B^2|H^{(\pm)}|^2 \notag\\
&+V_{UV}(H^{(\pm)})\delta(y)+V_{IR}(H^{(\pm)})\delta(y-L)\bigg\},    \label{action_SM}
\end{align}
where $F^a_{MN}$ and $B_{MN}$ are the 5D field strength tensors for $SU(2)$ and $U(1)_Y$, respectively with $a$
being the number generators of $SU(2)$. Above, $H^{(\pm)}$ are the even and odd $SU(2)$ doublets and the brane potentials are
\beq
V_{UV}(H^{(\pm)})=\frac{m^2_{UV}}{k}|H^{(\pm)}|^2,   \Lsp V_{IR}(H^{(\pm)})=-\frac{m^2_{IR}}{k}|H^{(\pm)}|^2+\frac{\lambda_{IR}}{k^2}|H^{(\pm)}|^4.
\label{boudary_potentials}
\eeq
In our approach, we do not put the Higgs quartic terms in the bulk nor on the UV-brane since we want EWSB to take place near the IR-brane. The covariant derivative $D_M$ is defined as follows:
\beq
D_M =\partial_M-i\frac{g_5}{2} \tau^aA^a_M -i\frac{g_5^\p}{2} B_M,      \label{co_dir_M_SM}
\eeq
where $\tau^a$ are Pauli matrices and $g_5(g^\p_5)$ is the coupling constant for the $A_M^a(B_M)$ fields.

We choose the 5D axial gauge by taking $V_5(x,y)=0$, where $V_5=W^\pm_5,Z_5,A_5$. Note that after choosing the 5D axial gauge there remains a 4D residual gauge
transformation $\widehat U(x)$ (independent of $y$), which is even under the geometric parity \cite{Ahmed:2015ona}. We rewrite the Higgs doublets in the following form:
\beq
\bpm H^{(+)} \\ H^{(-)}\epm=e^{ig_5 (\Pi^{(+)}\mathds{1}+\Pi^{(-)}\tau_1)} \bpm {\cal H}^{(+)} \\ {\cal H}^{(-)}\epm,   \label{higgs_redef}
\eeq
where $\mathds{1}$ and $\tau_1$ are the unit  and first Pauli matrices, respectively.
Above ${\cal H}$ and $\Pi$ are defined as (the parity indices are suppressed)
\begin{align}
{\cal H}(x,y)&\equiv \frac{1}{\sqrt2}\bpm 0\\ h(x,y)\epm,         \label{hat_H}\\
\Pi(x,y)&\equiv \bpm \frac{\cos^2\theta-\sin^2\theta}{2\cos\theta}\pi_Z & \frac{1}{\sqrt2}\pi^+_W\\
\frac{1}{\sqrt2}\pi^-_W& -\frac{1}{2\cos\theta}\pi_Z \epm, \lsp \text{where}\lsp \begin{array}{c}\cos\theta\equiv \frac{g_5}{\sqrt{g^2_5+g^{\p2}_5}} \\
\sin\theta\equiv \frac{g^\p_5}{\sqrt{g^2_5+g^{\p2}_5}} \end{array}.      \label{pi_matrix}
\end{align}

The KK-decomposition of the Higgs doublets $H^{(\pm)}(x,y)$ and the gauge fields $V^{(\pm)}_\mu(x,y)$ is
\begin{align}
{\cal H}^{(\pm)}(x,y)&=\sum_n {\cal H}^{(\pm)}_n(x)f^{(\pm)}_n(y),   \label{KK_H_sm}\\
\pi^{(\pm)}_{\tilde V}(x,y)&=\sum_n \pi_{{\tilde V}n}^{(\pm)}(x)a^{(\pm)}_{\tilde Vn}(y),   \label{KK_pi_sm}\\
V^{(\pm)}_\mu(x,y)&=\sum_n V^{(\pm)}_{\mu n}(x)a^{(\pm)}_{V_n}(y),   \label{KK_A_sm}
\end{align}
where $V_\mu=\tilde V_\mu(W^\pm_\mu,Z_\mu),A_\mu$. The wave-functions $f^{(\pm)}_n(y)$ and $a^{(\pm)}_{V_n}(y)$ satisfy
\begin{align}
-\partial_5 (e^{4A(y)}\partial_5 f^{(\pm)}_n(y))+\mu_B^2e^{4A(y)}f^{(\pm)}_n(y)&=m_n^{2(\pm)}e^{2A(y)}f^{(\pm)}_n(y),  \label{eom_fn_H}\\
-\partial_5 (e^{2A(y)}\partial_5 a^{(\pm)}_{V_n}(y))&=m_{V^{(\pm)}_n}^2a^{(\pm)}_{V_n}(y),  \label{eom_fn_A}\\
2\int_{0}^{L}dy e^{2A(y)}f^{(\pm)}_m(y)f^{(\pm)}_n(y)=\delta_{mn}, \lsp &2\int_{0}^{L}dy a^{(\pm)}_{V_m}(y)a^{(\pm)}_{V_n}(y)=\delta_{mn}, \label{norm_condition_HA_sm}
\end{align}
where the warped function $A(y)=-k|y|$.
The above KK-modes are subject to the following b.c.
\begin{align}
\left(\partial_5 -\frac{m^2_{UV}}{k}\right)f^{(+)}_n(y)\Big\vert_{0}&=0,    &f^{(-)}_n(y)\Big\vert_{0}&=0,  \label{H_bc0}\\
\partial_5a^{(+)}_{V_n}(y)\Big\vert_{0}&=0,    &a^{(-)}_{V_n}(y)\Big\vert_{0}&=0.  \label{A_bc0}\\
\left(\pm\partial_5-\frac{m^2_{IR}}{k}\right)f^{(\pm)}_{n}(y)\Big\vert_{L}&=0,  &\partial_5a^{(\pm)}_{V_n}(y)\Big\vert_{L}&=0.   \label{HA_bcL}
\end{align}

Under the assumption that the KK-scale is high enough, i.e.  $m_{KK}\sim\co(\text{few})\tev$,  we can consider
an effective theory where only the lowest modes (zero-modes with masses much below $m_{KK}$) are allowed.
It is important to note that the odd zero-mode wave functions obey  $a^{(-)}_{V_0}(y)=0$, as
can be easily seen from Eq. \eqref{eom_fn_A} along with the b.c. \eqref{A_bc0} and \eqref{HA_bcL}.
As a consequence of $a^{(-)}_{V_0}(y)=0$,   the odd zero-mode gauge fields $V^{(-)}_{0\mu}(x)$
and the odd Goldstone modes $\pi^{(-)}_{\tilde V_0}(x)$ will not be present in the effective 4D theory. Moreover, the even zero-mode gauge
profile is constant, i.e.  $a^{(+)}_{V_0}(y)=1/\sqrt{2L}$. To proceed, we introduce a convenient notion for our zero-mode effective theory by redefining $V_{0\mu}^{(+)}(x)\equiv V_\mu(x)$, $\pi^{(+)}_{\tilde V0}(x)\equiv \pi_{\tilde V}(x)$, $\Pi^{(+)}_{0}(x)\equiv \widehat \Pi(x)$ and
\beq
H_1(x)\equiv e^{ig_4 \widehat \Pi(x)}{\cal H}^{(+)}_0(x), \lsp H_2(x)\equiv e^{ig_4 \widehat \Pi(x)}{\cal H}^{(-)}_0(x).  \label{H_12}
\eeq
The zero-mode effective action can be written as
\begin{align}
S_{eff}=-\int d^4x &\bigg\{\frac14 F_{\mu\nu}{ F}^{\mu\nu}+\frac14 { Z}_{\mu\nu}{Z}^{\mu\nu}+\frac12 { W}^{+}_{\mu\nu}{W}^{-\mu\nu} \notag\\
&+ \big({\cal D}_\mu H_1\big)^{\dag}{\cal D}^\mu H_1+ \big({\cal D}_\mu H_2\big)^{\dag}{\cal D}^\mu H_2 +V(H_1,H_2) \bigg\},    \label{eff_action_nab_gauge_cov}
\end{align}
where the scalar potential can be written as
\begin{align}
V(H_1,H_2) =&-\mu^{2}|H_1|^2 -\mu^{2}|H_2|^2 +\lambda|H_1|^4 +\lambda|H_2|^4+6\lambda|H_1|^2|H_2|^2.    \label{potenial_sm}
\end{align}
The covariant derivative ${\cal D}_\mu$ is defined as
\begin{align}
{\cal D}_\mu&=\partial_\mu-ig_4\mathbb{\hat A}_{\mu}, \Lsp g_4(g_4^\p)\equiv \frac{g_5(g_5^\p)}{\sqrt 2L},  \label{co_dir_4_sm}
\end{align}
where $\mathbb{\hat A}_{\mu}$ is defined as
\begin{align}
\mathbb{\hat A}_{\mu}(x)&\equiv \bpm \sin\theta A_{\mu}+\frac{\cos^2\theta-\sin^2\theta}{2\cos\theta}Z_{\mu} & \frac{1}{\sqrt2}W^+_{\mu}\\
\frac{1}{\sqrt2}W^-_{\mu}& -\frac{1}{2\cos\theta}Z_{\mu} \epm.      \label{A_matrix_0}
\end{align}
In the above scalar potential the mass parameter $\mu$ and quartic coupling $\lambda$ are defined as,
\beq
\mu^2\equiv-m_{0}^{2(\pm)}=(1+\beta)m_{KK}^2\delta_{IR}, \Lsp \lambda\equiv\lambda_{IR}(1+\beta)^2, 	\label{mu_lambda}
\eeq
where $\delta_{IR}$, $m_{KK}$ and $\beta$ are given by
\beq
\delta_{IR}\equiv\frac{m^2_{IR}}{k^2}-2(2+\beta),  \lsp m_{KK}\equiv ke^{-kL} \hsp\text{and}\hsp\beta\equiv\sqrt{4+\mu^2_B/k^2}.
\label{delta_IR_mKK}
\eeq

Concerning the symmetries of the above potential, one can notice that $V(H_1,H_2)$ is invariant under $[SU(2)\times U(1)_Y]^\prime \times [SU(2)\times U(1)_Y]$, where one of the blocks has been gauged while the other one survived as a global symmetry.
The zero-modes of the four odd vector bosons $(W_{0\mu}^{(-)\pm}, Z_{0\mu}^{(-)} \text{ and } A_{0\mu}^{(-)})$ and the three would-be-Goldstone
bosons $\Pi^{(-)}_0$ have been removed by appropriate b.c., implying that the
corresponding gauge symmetry has been broken explicitly. What remains is {\it the
truncated inert doublet model}, that contains ${H}_{1,2}$, and the corresponding residual $SU(2)\times U(1)_Y$ global symmetry of the action.
Symmetry under the above mentioned $U(1)^\prime \times U(1)$ implies in particular that
$V(H_1,H_2)$ is also invariant under various $\mathbb{Z}_2$'s, for example
$H_1\to -H_1$, $H_2\to -H_2$ and $H_1\to \pm H_2$.

\begin{figure}[t]
\begin{center}
\includegraphics[scale=0.45]{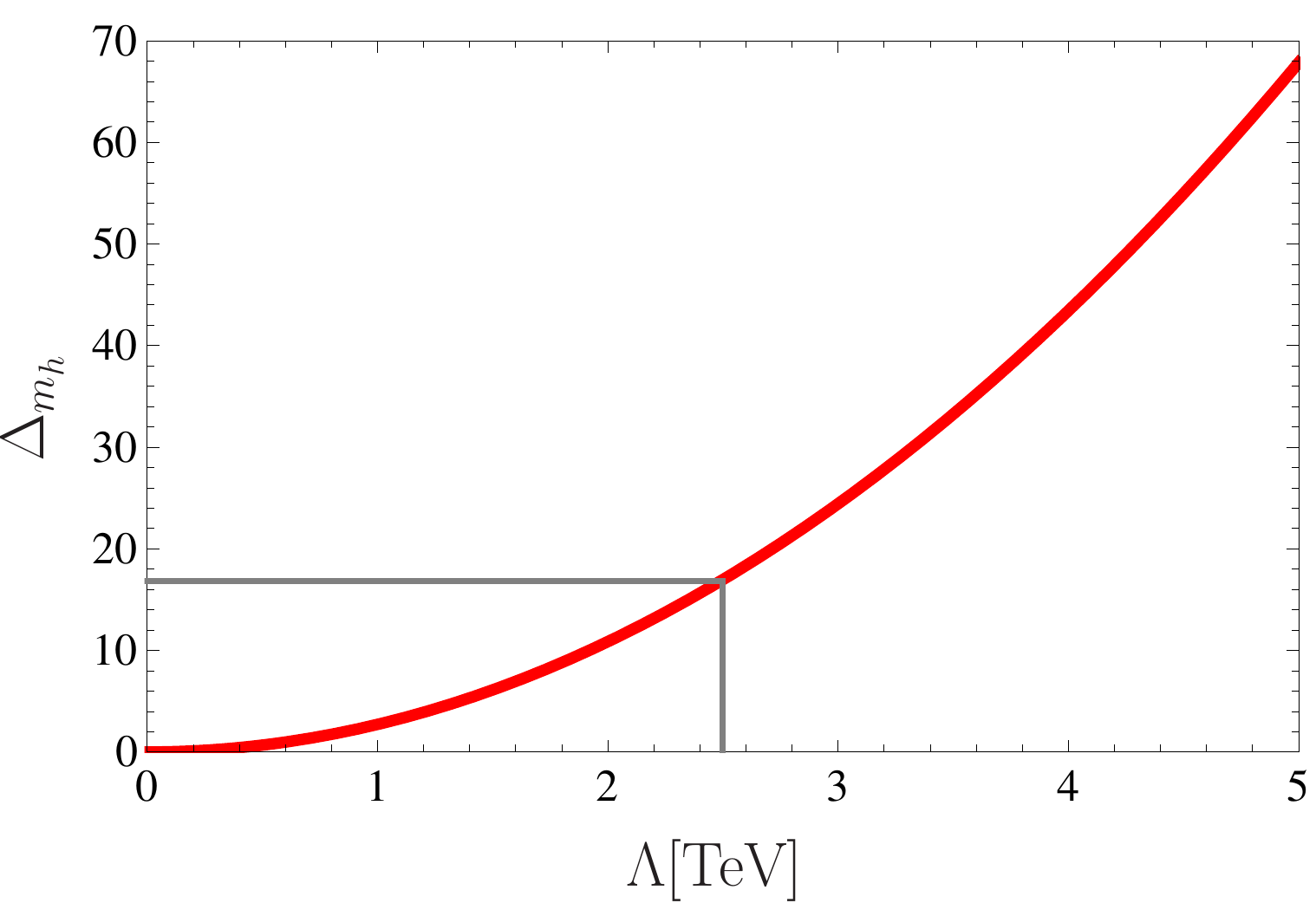}\hsp
\includegraphics[scale=0.45]{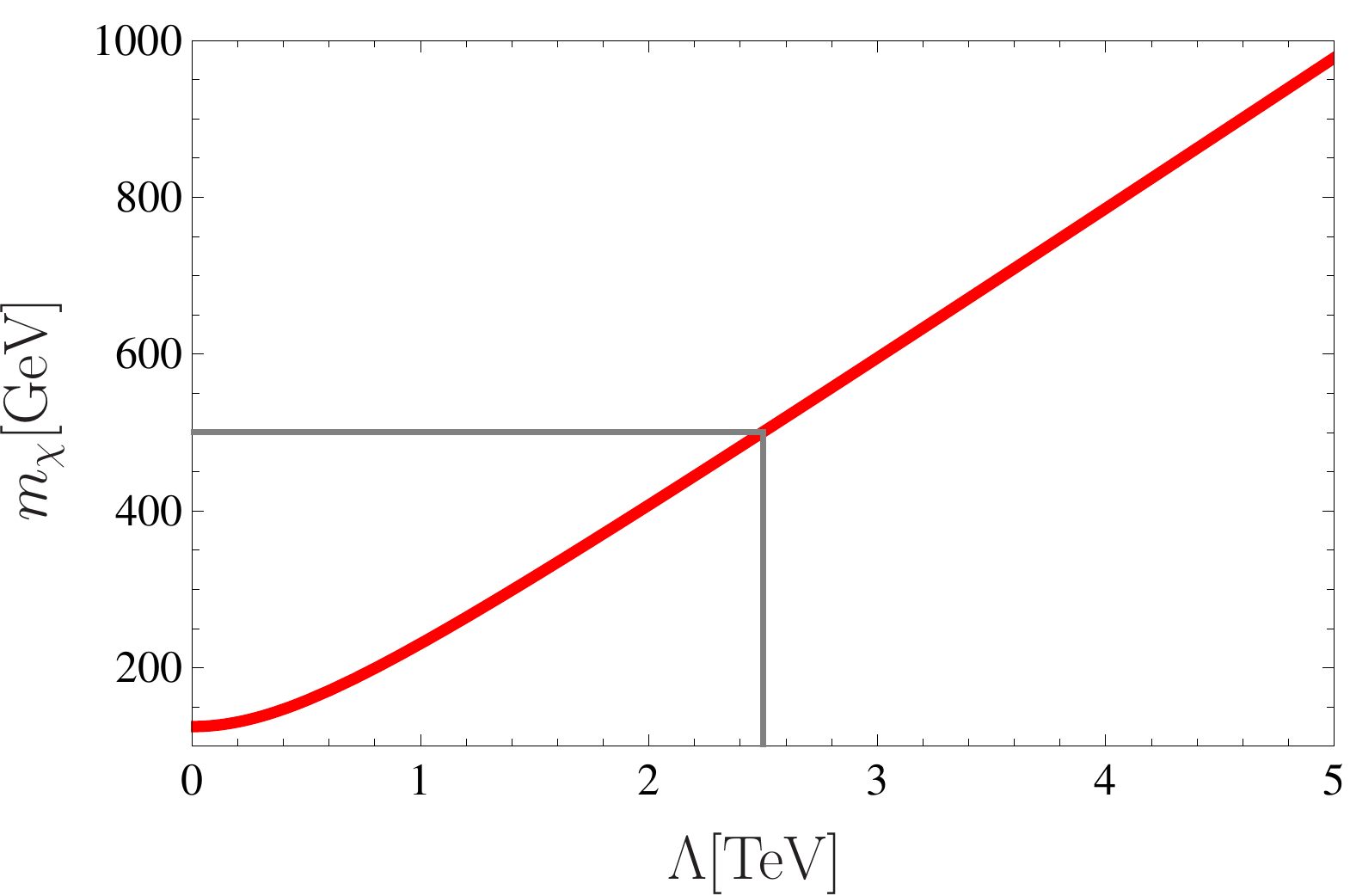}
\end{center}
\caption{The left plot gives the value of the fine-tuning measure $\Delta_{m_h}$ for a Higgs mass of $125$ GeV as a function of the cutoff $\Lambda$.  The  right plot shows the dependence of $m_\chi$ on $\Lambda$ for $m_h=125\gev$. In our model $\Lambda=m_{KK}$.  The vertical gray line indicates the  current lower bound on the KK mass scale coming from  EWPT as computed in our model for $\beta=0$,    $m_{KK}\gtrsim2.5\tev$.}
\label{mchi_lambda}
\end{figure}

The above potential has four degenerate vacua \cite{Ahmed:2015ona}, we choose the vacuum such that the Higgs field $H_1$ acquires a vev,
whereas the Higgs field $H_2$ does not, i.e.
\beq
v^2_1\equiv v^2=\frac{\mu^2}{\lambda},  \Lsp v_2=0.        \label{v1_v2}
\eeq
Fluctuations around the vacuum of our choice are
\beq
H_1(x)=\frac{1}{\sqrt2}e^{ig_4\widehat \Pi}\bpm 0\\v+h\epm,    \Lsp H_2(x)=\frac{1}{\sqrt2}e^{ig_4\widehat \Pi}\bpm 0\\ \chi \epm,    \label{H1_H2_def_sm}
\eeq
where $\widehat \Pi$ contains the pseudoscalar Goldstone bosons $\pi_{W^\pm,Z}$. We choose the unitary gauge in which $\pi_{W^\pm,Z}$ are gauged away and the gauge bosons $W^\pm_\mu$ and $Z_\mu$ become massive.
In the unitary gauge our effective action is
\begin{align}
S_{eff}=-\int d^4x &\bigg\{\frac12 {\cal W}^{+}_{\mu\nu}{\cal W}^{-\mu\nu}+ \frac14 {\cal Z}_{\mu\nu}{\cal Z}^{\mu\nu}+\frac14 {\cal F}_{\mu\nu}{\cal F}^{\mu\nu}+ m^2_W W^+_\mu W^{-\mu}+\frac12 m^2_Z Z_\mu Z^\mu\notag\\
&+\frac12\partial_\mu h\partial^\mu h+ \frac12m^2_h h^2  +\frac12\partial_\mu \chi\partial^\mu \chi+ \frac12m^2_\chi \chi^2+\lambda vh^3+\frac\lambda4 h^4 +\frac\lambda4 \chi^4 \notag\\
&+3\lambda v h\chi^2+\frac32\lambda h^2\chi^2+\frac{g_4^2}{2}vW_\mu^+W^{-\mu}h +\frac{g_4^2}{4}W_\mu^+W^{-\mu}(h^2+\chi^2)  \notag\\
&+\frac14(g_4^2+g_4^{\p2})vhZ_\mu Z^\mu  +\frac18(g_4^2+g_4^{\p2})Z_\mu Z^\mu (h^2+\chi^2)\bigg\},    \label{eff_action_quadratic}
\end{align}
where the masses are,
\begin{align}
m^2_h&=2\mu^2, \lsp m^2_\chi=2\mu^2+\frac34\frac{\Lambda^2}{\pi^2v^2}m_t^2, \lsp m^2_{W}=\frac{g_4^2m^2_{Z}}{g^2_4+g^{\p2}_4}=\frac{1}{4}g^2_4\frac{\mu^2}{\lambda}.         \label{masses_higgs_WZ}
\end{align}
It is worth noticing here that the Higgs mass $m_h$ and the dark scalar mass $m_\chi$ are modified by quantum corrections. The above masses are those obtained after taking into account 1-loop quadratically divergent contributions within the effective theory. However, to get the Higgs mass $m_h=125\gev$ we need to fine-tune the parameters of the theory. We plot the fine-tuning measure $\Delta_{m_h}$ (defined in Ref.~\cite{Ahmed:2015ona}) as a function of the effective cutoff scale $\Lambda\equiv m_{KK}$ in Fig. \ref{mchi_lambda}. The most stringent bounds on the KK-scale $m_{KK}$ in the RS1 geometry with a bulk Higgs come from  electroweak precision tests (EWPT) by fitting the $S,~T$ and $U$ parameters \cite{Archer:2014jca,Dillon:2014zea}. The lower bound on the KK mass scale in our model (AdS geometry,    i.e.  $A(y)=-k|y|$) is $m_{KK}\gtrsim2.5\tev$ for $\beta=0$ and $m_{KK}\gtrsim4.3\tev$ for $\beta=10$ at $95\%$ C.L. \cite{Archer:2014jca}. This implies a tension between fine-tuning (naturalness) and the lower bound on the KK mass scale $m_{KK}$. The region within the gray lines in Fig. \ref{mchi_lambda} shows the current bounds on the KK mass scale for our geometry and the associated fine-tuning.

As illustrated in Fig.~\ref{mchi_lambda} the dark matter mass $m_\chi$ is raised linearly with the cut-off scale $\Lambda$. An interesting aspect of our model is that dark matter is predicted to be heavier than the SM Higgs boson.
A natural value of the cutoff coincides with the mass of the first KK excitations, which are experimentally
limited to lie above a few TeV (depending on model details and KK mode sought). The strongest version of the EWPT bound requires $m_{KK}\gsim 2.5\tev$ \cite{Archer:2014jca}, corresponding to $m_\chi\gsim 500\gev$, for which $\Delta_{m_h}$ is a very modest $\sim 18$. Our model is most consistent for $500\gev\lsim m_\chi \lsim 1200\gev$, where the upper bound is placed by requiring that the fine-tuning measure $\Delta_{m_h}$ be less than $100$.

\subsection{Dark matter relic abundance}
\label{Dark matter relic abundance}
\noindent In this subsection we calculate the dark matter relic abundance. The diagrams contributing to dark matter annihilation are shown in Fig.~\ref{DM_annihi_diagrams}. The squared amplitudes $|{\cal M}|^2$ corresponding to the contribution of each final state  to dark matter annihilation are:
\begin{align}
\left| \mathcal{M} (\chi\chi \to \tilde V\tilde V)\right|^2 &= {4 m_{\tilde V}^4 \over S_{\tilde V} v^4} \left( 1+ {3m_h^2 \over s-m_h^2} \right)^2 \left[ 2+ \left( 1- {s \over 2m^2_{\tilde V}} \right)^2 \right],  \label{M2VV}\\
\left| \mathcal{M} (\chi\chi \to f\bar{f}) \right| ^2&=  18 N_c {m_f^2 m_h^4 \over v^4} {s-4m^2_f \over (s-m_h^2)^2}, \label{M2ff}\\
\left| \mathcal{M} (\chi\chi \to hh) \right|^2 &= {9 m_h^4 \over 2 v^4} \left[ 1+ 3m_h^2 \left( {1 \over s-m_h^2} + {1 \over t-m_\chi^2} + {1 \over u-m_\chi^2} \right) \right]^2,  \label{M2hh}
\end{align}
where $\tilde V={W,Z}$ and $S_W=1$ and $S_Z=2$ are symmetry factors accounting for identical particles in the final state; $N_c$ refers to the number of ``color'' degrees of freedom for the given fermion and $s,~t,~u$ are the Mandelstam variables. Here, we ignore the loop-induced $\gamma\gamma$ and $Z\gamma$ final states, which are strongly suppressed.
Note that the first term in the parenthesis in Eq.~(\ref{M2VV}) and the first term in the square bracket in Eq.~(\ref{M2hh}) arise from the $\chi\chi \tilde V\tilde V$ and the $\chi\chi hh$ contact interactions, respectively.
The former channel is present in our model since $\chi$ is a component of the  (truncated) odd $SU(2)$ doublet.
\begin{figure}[t]
\centering
\begin{tikzpicture}[node distance=1cm,very thick, rounded corners=0pt,line cap=round]
\begin{scope}[xshift=0.5cm]
\coordinate[] (v1);
\coordinate[above left=of v1] (a1);
\coordinate[below left=of v1] (a2);
\coordinate[above right=of v1] (b1);
\coordinate[below right=of v1] (b2);
\draw[scalar] (a1)node[left]{$\chi$} -- (v1);
\draw[scalar] (a2)node[left]{$\chi$} -- (v1);
\draw[boson](v1)--(b1)node[right]{$W,Z$};
\draw[boson](v1)--(b2)node[right]{$W,Z$};
\filldraw [blue] (v1) circle (2pt);
\end{scope}
\begin{scope}[xshift=4.5cm]
\coordinate[] (v1);
\coordinate[above left=of v1] (a1);
\coordinate[below left=of v1] (a2);
\coordinate[right=1.3cm of v1] (v2);
\coordinate[above right=of v2] (b1);
\coordinate[below right=of v2] (b2);
\draw[scalar] (a1)node[left]{$\chi$} -- (v1);
\draw[scalar] (a2)node[left]{$\chi$} -- (v1);
\draw[scalar] (v1)--node[above]{$h$}(v2);
\draw[boson](v2)--(b1)node[right]{$W,Z$};
\draw[boson](v2)--(b2)node[right]{$W,Z$};
\filldraw [blue] (v1) circle (2pt)
                (v2) circle (2pt);
\end{scope}
\begin{scope}[xshift=9.5cm]
\coordinate[] (v1);
\coordinate[above left=of v1] (a1);
\coordinate[below left=of v1] (a2);
\coordinate[right=1.3cm of v1] (v2);
\coordinate[above right=of v2] (b1);
\coordinate[below right=of v2] (b2);
\draw[scalar] (a1)node[left]{$\chi$} -- (v1);
\draw[scalar] (a2)node[left]{$\chi$} -- (v1);
\draw[scalar] (v1)--node[above]{$h$}(v2);
\draw[fermion](v2)--(b1)node[right]{$f$};
\draw[fermion](b2)node[right]{$\bar f$} -- (v2);
\filldraw [blue] (v1) circle (2pt)
                (v2) circle (2pt);
\end{scope}\newline
\begin{scope}[yshift=-2.5cm]
\coordinate[] (v1);
\coordinate[above left=of v1] (a1);
\coordinate[below left=of v1] (a2);
\coordinate[above right=of v1] (b1);
\coordinate[below right=of v1] (b2);
\draw[scalar] (a1)node[left]{$\chi$} -- (v1);
\draw[scalar] (a2)node[left]{$\chi$} -- (v1);
\draw[scalar](v1)--(b1)node[right]{$h$};
\draw[scalar](v1)--(b2)node[right]{$h$};
\filldraw [blue] (v1) circle (2pt);
\end{scope}
\begin{scope}[xshift=3cm,yshift=-2.5cm]
\coordinate[] (v1);
\coordinate[above left=of v1] (a1);
\coordinate[below left=of v1] (a2);
\coordinate[right=1.3cm of v1] (v2);
\coordinate[above right=of v2] (b1);
\coordinate[below right=of v2] (b2);
\draw[scalar] (a1)node[left]{$\chi$} -- (v1);
\draw[scalar] (a2)node[left]{$\chi$} -- (v1);
\draw[scalar] (v1)--node[above]{$h$}(v2);
\draw[scalar](v2)--(b1)node[right]{$h$};
\draw[scalar](v2) -- (b2)node[right]{$h$};
\filldraw [blue] (v1) circle (2pt)
                (v2) circle (2pt);
\end{scope}
\begin{scope}[xshift=7.5cm,yshift=-2cm]
\coordinate[] (v1);
\coordinate[below= of v1] (v2);
\coordinate[position=150 degrees from v1] (a1);
\coordinate[position=-150 degrees from v2] (a2);
\coordinate[position=30 degrees from v1] (b1);
\coordinate[position=-30 degrees from v2] (b2);
\draw[scalar] (a1)node[left]{$\chi$} -- (v1);
\draw[scalar] (a2)node[left]{$\chi$} -- (v2);
\draw[scalar] (v1)--node[left]{$\chi$}(v2);
\draw[scalar](v1)--(b1)node[right]{$h$};
\draw[scalar](v2)--(b2)node[right]{$h$};
\filldraw [blue] (v1) circle (2pt)
                (v2) circle (2pt);
\end{scope}
\begin{scope}[xshift=11cm,yshift=-2cm]
\coordinate[] (v1);
\coordinate[below= of v1] (v2);
\coordinate[position=150 degrees from v1] (a1);
\coordinate[position=-150 degrees from v2] (a2);
\coordinate[above right=1.8cm of v2] (b1);
\coordinate[below right=1.8cm of v1] (b2);
\draw[scalar] (a1)node[left]{$\chi$} -- (v1);
\draw[scalar] (a2)node[left]{$\chi$} -- (v2);
\draw[scalar] (v1)--node[left]{$\chi$}(v2);
\draw[scalar](v2)--(b1)node[right]{$h$};
\draw[scalar](v1)--(b2)node[right]{$h$};
\filldraw [blue] (v1) circle (2pt)
                (v2) circle (2pt);
\end{scope}
\end{tikzpicture}
\caption{Dark matter annihilation diagrams.}
\label{DM_annihi_diagrams}
\end{figure}
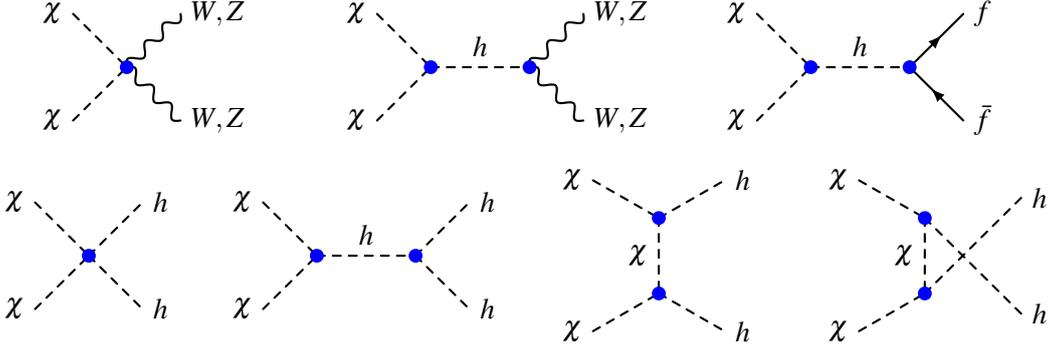
\begin{figure}[t]
\begin{center}
\includegraphics[width=0.47\textwidth]{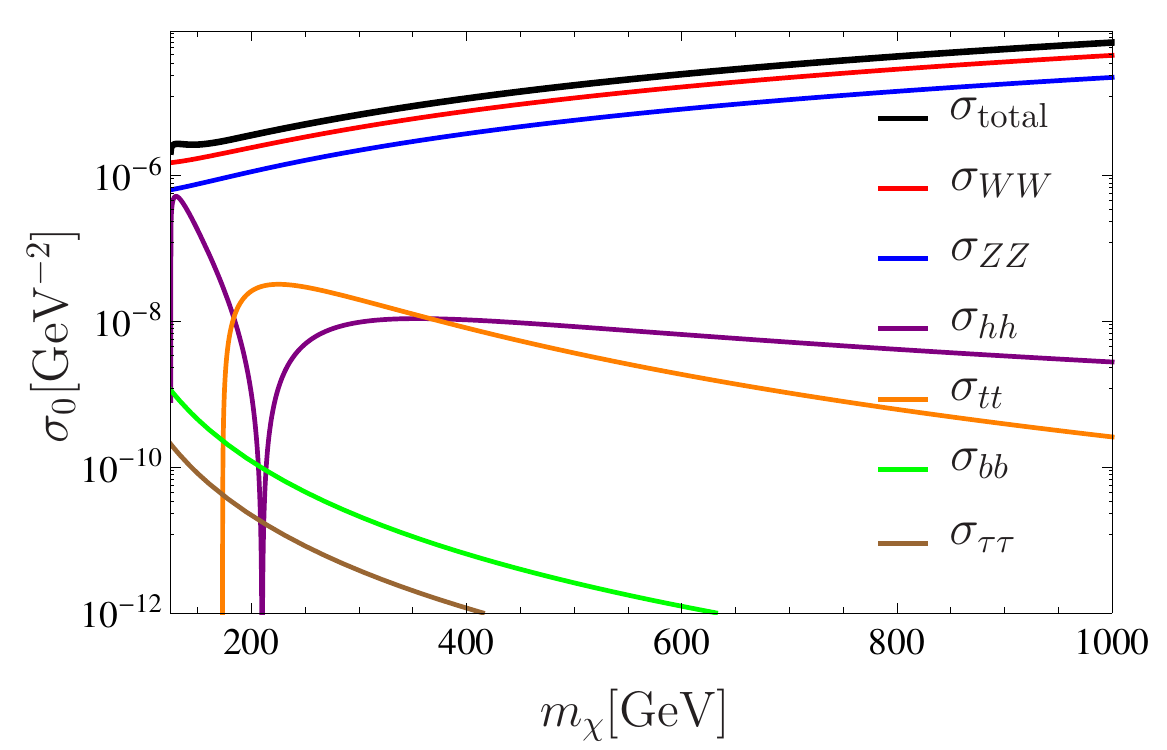}\hsp
\includegraphics[width=0.45\textwidth]{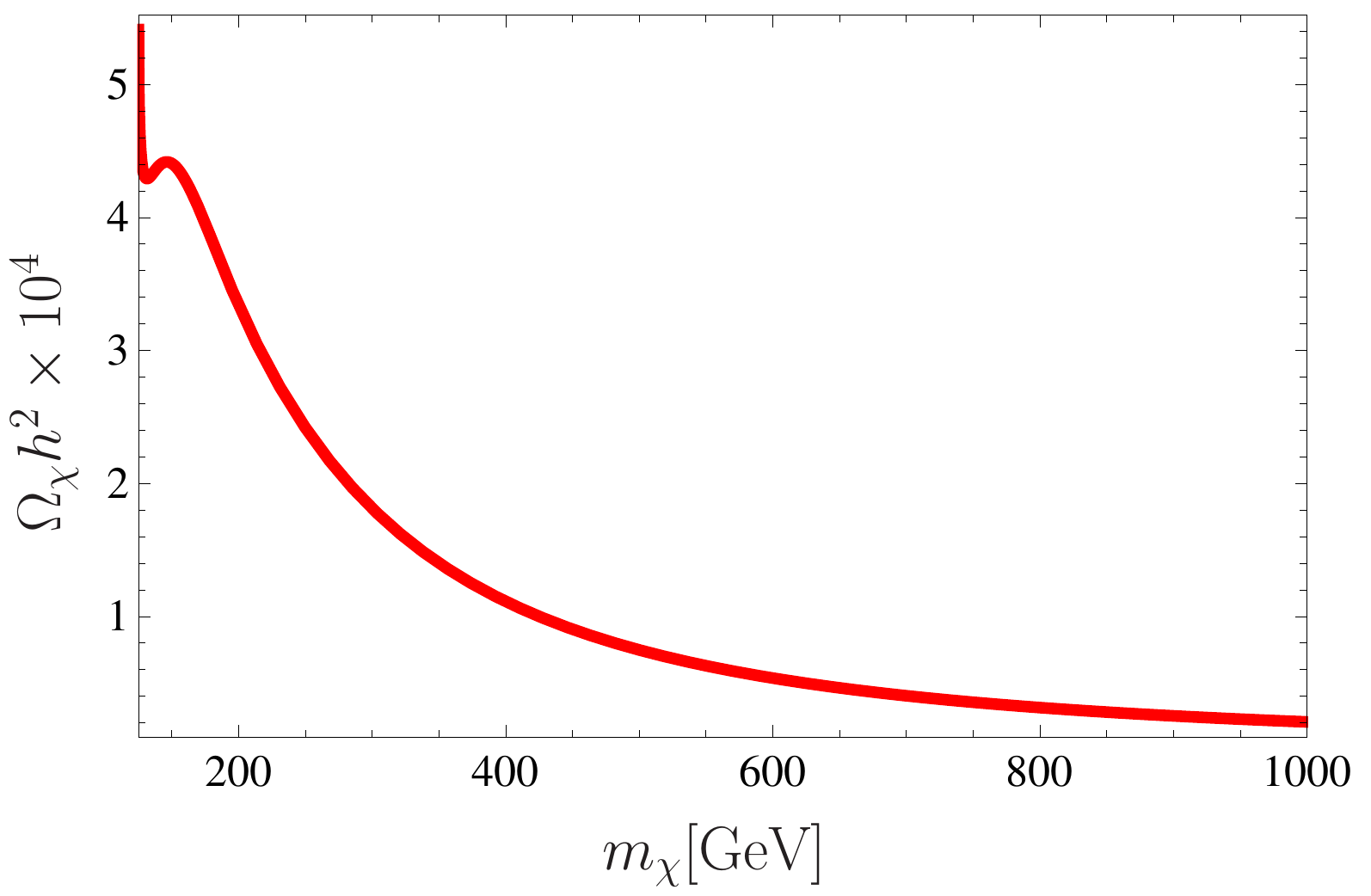}
\end{center}
\caption{The above graphs show the annihilation cross-section $\sigma_0$ for different final states (left)
and the $\chi$ abundance $\Omega_\chi h^2\times 10^{4}$ (right) as a function of dark matter mass $m_\chi$.}
\label{sigma_omega}
\end{figure}

In Fig.~\ref{sigma_omega} (left panel) we have plotted the annihilation cross-section for
the contributing channels as a function of $m_\chi$. As shown in the graph the total cross section is dominated by $WW$ and $ZZ$ final states. The main contributions for these final states are those generated by contact interactions $\chi\chi WW(ZZ)$, whereas, all the other final states that include the Higgs boson $h$ or the top quark are very small in comparison to $\chi\chi \to WW(ZZ)$. The dark matter relic abundance $\Omega_\chi h^2$ is shown in Fig.~\ref{sigma_omega} (right panel). We observe that $\Omega_\chi h^2\lsim 10^{-4} $ once the electroweak precision bound on the KK mass scale $m_{KK}$ is imposed \cite{Ahmed:2015ona}.

\section{Conclusions}
\label{Summary}
\noindent In this article, we constructed a model with ${\mathbb Z}_2$ geometric symmetry which allows a warped KK-parity such that all the bulk fields are either even or odd under this parity.  We employed the SM gauge sector in the bulk of the $Z_2$-symmetric geometry and analysed EWSB due to the bulk Higgs. The zero-mode effective theory appropriate at scales below the KK scale,  $m_{KK}$, was obtained. The resulting model has the following features.
\ben \itemsep0em
\item In the $\mathbb{Z}_2$-symmetric IR-UV-IR model, due to warped KK-parity all the bulk fields develop even and odd towers of KK-modes in the 4D effective theory.
\item We have shown that the physics of full IR-UV-IR model is equivalent to two times the RS1 geometry (UV-IR) provided all the bulk fields are subject to the Neumann (or mixed) and Dirichlet b.c corresponding to even and odd parity fields.
\item Assuming that the KK-scale is high enough ($m_{KK}\sim\co(\text{few})\tev$), we have derived the low energy effective theory which includes only zero-modes of the theory.
\item In the low energy (zero-mode) effective theory, we have all the SM fields plus a {\it dark-Higgs} -- dark matter candidate. In the low energy effective theory we have calculated the mass of the SM Higgs and the dark-Higgs. Using the strongest version of the EWPT bound $m_{KK}\gsim 2.5\tev$, one gets the lower bound on the dark-Higgs mass to be $500\gev$. In the end, our model is most consistent for $500\gev\lsim m_\chi \lsim 1200\gev$ if we allow a maximum of $1\%$ fine-tuning in the parameters of the theory.
\item We have calculated the relic abundance of the dark-Higgs in the cold dark matter approximation. For $m_\chi$ in the above preferred range,  $\Omega_\chi h^2 \lsim 10^{-4}$ as compared to the current experimental value of $\sim 0.1$.
\een

\section*{Acknowledgements}
\noindent The work of AA and BG has been supported in part by the National Science Centre
(Poland) as  research projects no DEC-2014/15/B/ST2/00108
and DEC-2014/13/B/ST2/03969. JFG and YJ are supported in part by US DOE grant DE-SC-000999.


\providecommand{\href}[2]{#2}\begingroup\raggedright\endgroup


\end{document}